\documentclass{article}
\usepackage{frascatiphys}
\usepackage{amsmath}
\usepackage{relsize}
\usepackage{wasysym}
\usepackage{amssymb}
\usepackage{graphicx}
\newcommand{\be}{\begin{equation}}
\newcommand{\ee}{\end{equation}}
\newcommand{\ba}{\begin{array}}
\newcommand{\ea}{\end{array}}
\newcommand{\bea}{\begin{eqnarray}}
\newcommand{\eea}{\end{eqnarray}}
\newcommand{\sss}{\scriptscriptstyle}

\newcommand{\nn}{\nonumber}
\renewcommand{\L}{{\sss L}}

\newcommand{\hc}{{\sss HC}}
\begin{document}
\title{ 
$B$ decay anomalies and dark matter from strong dynamics
}
\author{
James M.\ Cline\\
Niels Bohr International Academy, Copenhagen, Denmark,\\ and\\
McGill University, Physics Department, Montr\'eal, Canada
}
\maketitle
\baselineskip=10pt
\begin{abstract}
Indications of lepton flavor universality violation in semileptonic
$B$ decays to $K$ or $K^*$ and muons or electrons can be explained
by leptoquark exchange.  I present a model in which the leptoquark
is a bound state of constituents charged under a new confining
SU($N_{\rm HC}$) hypercolor interaction.  The lightest neutral bound
state in the theory is an asymmetric dark matter candidate, that might
be directly detectable through its magnetic dipole moment interaction. 
\end{abstract}
\baselineskip=14pt

\section{Introduction} Strong dynamics has been a useful idea for
going beyond the standard model (SM) in the context of several tentative
experimental anomalies from the past, such as the 750 GeV diphoton
excess at LHC and the 130 GeV gamma ray excess at the Fermi telescope.
It has also proved useful for building models of composite dark matter 
arising from a possibly rich hidden sector.  One motivation for such
models is the hint of strong dark matter self-interactions from 
cosmological $N$-body simulations versus observations of galactic
structure.

Recently the LHCb experiment at CERN has found tentative evidence for
violation of lepton flavor universality in the decays of $B\to K$ or
$K^*$ and $e^+e^-$ or $\mu^+\mu^-$ \cite{Aaij:2014ora,Aaij:2017vbb}.
Popular theoretical explanations
involve tree-level exchange of new heavy states---gauge bosons or
leptoquarks, or exchange of new particles in a loop.  Here I will
focus on the leptoquark option.  One might consider the leptoquark as
a rather odd beast in the beyond-the-standard-model zoo, not being
required by any new principles such as supersymmetry.  However in the
context of strong dynamics the existence of leptoquarks can be
natural, since all that is required is that new hypercolored fields 
interacting with quarks or leptons respectively become bound to each
other in a meson-like state.  If there also exist new fields that are
neutral under the standard model symmetries but charged under
hypercolor, then a composite dark matter candidate comes at no extra
cost, beyond considerations of its stability.

Model-independent fits, for example refs.\ \cite{DAmico:2017mtc,Hiller:2017bzc}, show that the new physics can be well-described
by the addition of a single operator
\be
	{\cal O}_{b_\L\mu_\L} = 
{c\over\Lambda^2}(\bar s_\L\gamma_\alpha b_\L)(\bar\mu_\L\gamma^\alpha \mu_\L)
\label{effop}
\ee
in the effective Hamiltonian, with
\be
	{c\over\Lambda^2} = {1.0\times 10^{-3}\over{\rm TeV}^2}\,.
\ee
By a Fierz transformation, the operator (\ref{effop}) can be put into
the form more suggestive of leptoquark exchange, 
$(\bar s_\L\gamma_\alpha \mu_\L)(\bar\mu_\L\gamma^\alpha s_\L)$.

\section{Model}
Our model \cite{Cline:2017aed} introduces three new particles: 
a vectorlike quark partner $\Psi$ and right-handed 
neutrino partner $S$, and an inert Higgs doublet $\phi$, 
all charged under SU(N)$_\hc$ and an accidental $Z_2$, which are
listed in table \ref{t1}.
The allowed couplings to standard model 
left-handed quarks and leptons are
\be
{\cal L} = 	\tilde\lambda_f\, \bar Q_{f,a}\, \phi^a_{A}
\Psi^{A} + 
	\lambda_f\, \bar S_{A}{\phi_{a}^{\!\!\!*A}}\, L_f^a
\label{Lint}
\ee
with $f$ being the generation index.  We work in a basis where the
mass matrices of the charged leptons and down-like quarks are presumed
to be diagonal, hence CKM mixing comes exclusively from
diagonalization of the up-like quark mass matrix.  After going to the
mass basis, the couplings to down-like quarks remain $\lambda_f$, but
those to up-like quarks are rotated, 
$\tilde\lambda_i \bar Q_i \to 
	\tilde \lambda_j\left(\begin{array}{cc} \bar u_{\L,i}
	V_{ij},\ 
	\bar d_{\L,j} \end{array}\right)
	\equiv \left(\begin{array}{cc} \tilde\lambda'_i \bar u_i,\ 
	 \tilde\lambda_i \bar d_i \end{array}\right)\,.
$

\begin{table}[t]
\centering
\begin{tabular}{|c|c|c|c|c|c|c|}
\hline
 & SU(3) & SU(2)$_L$ & U(1)$_y$ & U(1)$_{\rm em}$ & SU(N)$_\hc$ & $Z_2$\\
\hline
$\Psi$ & $3$ & $1$ & $2/3$ & $2/3$ & $N$ & $-1$\\
$S$    & $1$ & $1$ & $0$ & $0$ & $N$ & $-1$\\
$\phi$ & $1$ & $2$ & $-1/2$ & $(0,-1)$ &  $\bar N$ & $-1$\\
\hline
\end{tabular}
\caption{ New particles and their quantum numbers} 
 \label{t1}
\end{table}

Because of confinement by the hypercolor interaction, there are
various meson-like bound states $\bar S S$, $\bar\Psi\Psi$ and
$\bar\Psi S$, the last of which has leptoquark quantum numbers.
All of these states can have either spin 0 or spin 1.  The spin-0
(pseudoscalar) 
leptoquark $\Pi$ couples to SM fermions through a derivative interaction
since the matrix element $\langle 0|(\bar S \gamma_\mu\gamma_5\Psi)|\Pi\rangle = 
f_\Pi\, p_\Pi^\mu$ is analogous to that of the pion in QCD.  When 
$p_\Pi^\mu$ is contracted with the $\bar q\gamma_\mu \ell$ current of
the SM fermions, it leads to the small masses $m_q$ and $m_\ell$
following from the Dirac equation, which suppresses the matrix
element.  For this reason the vector leptoquark $\Phi_\mu$ interacts
more strongly with the SM fermions.  Its matrix element is
$\langle 0|(\bar S \gamma_\mu\Psi)|\Phi_\lambda\rangle =
	f_{\Phi}\,m_{\Phi}\epsilon_\lambda^\mu$ for a state with 
polarization labeled by $\lambda$.  

To determine the effective coupling $g_{\Phi}^{fg}$ of $\Phi_\mu$ to the SM fermions,
we can compare the decay rate computed in the effective theory,
$\Gamma(\Phi_\mu\to L_g\bar Q_f) =  {|g_\Phi^{fg}|^2\over 24\pi }m_\Phi$,
to its prediction in terms of the constituents in the bound state
\cite{Kang:2008ea},
\be
	\Gamma(\Phi_\mu\to L_g\bar Q_f) = 
	\sigma v_{\rm rel}(S\bar\Psi\to L_g\bar Q_f)
	|\psi(0)|^2
\ee
where $\psi(0)$ is the wave function of the bound state evaluated at
the origin, and $\sigma$ is the perturbative cross section for the
indicated scattering.  This gives
\be
	g_{\Phi}^{fg} = \left(N_\hc\over  4 m_\Phi\right)^{1/2}
{\tilde\lambda_f\lambda_g 
	\,(m_S + m_\Psi)\,\psi(0)\over
	(m_\phi^2+m_s m_\Psi)}
\ee

\begin{figure}[t]
\hspace{-0.4cm}
\centerline{
\includegraphics[width=0.5\hsize]{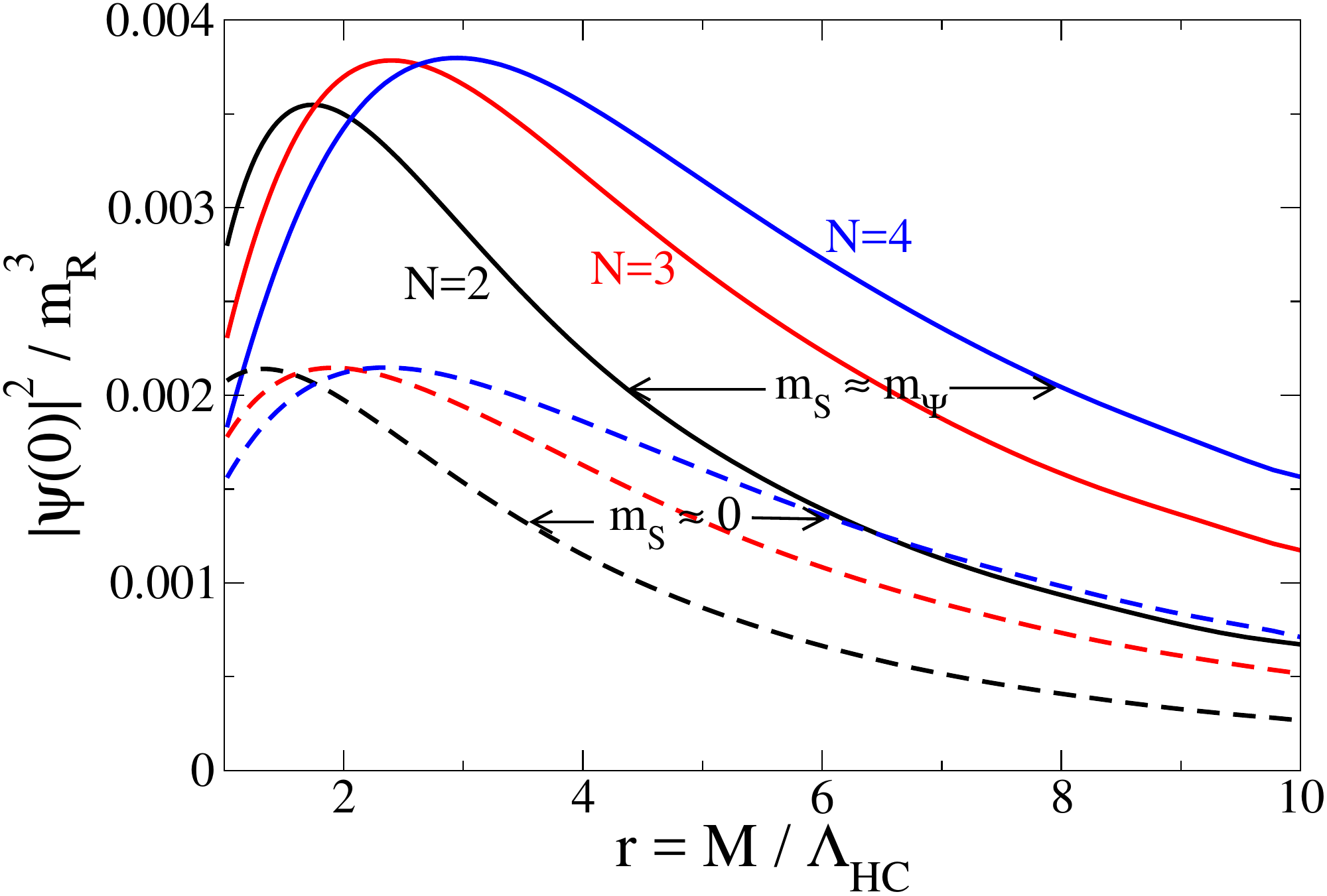}}
\caption{The function $\zeta = |\psi(0)|^2/m_\Phi^3$ (eq.\ (\ref{zeq}))
Solid curves correspond to 
both constituents (and the inert doublet $\phi$) having the same mass $M$, while dashed 
ones show the
case of $m_S\ll M$.}
\label{zetaf}
\end{figure}

We still need to determine $\psi(0)$.  We will be interested in heavy
constituent masses, of order the confinement scale $\Lambda_\hc$,
for which a nonrelativistic potential model should give reasonable
estimates.  We take a Cornell potential
\be
	V_c = -{\alpha_\hc\over 2 r}\left(N_\hc - {1\over N_\hc}\right)
	 +2(N_\hc-1)\Lambda_\hc^2\, r
\ee
between fundamental and anti-fundamental states, and a hydrogen-like
variational ansatz for the wave function, $\psi \sim e^{-\mu_*r/2}$.  The scale
$\mu_*$ and the mass of the bound state are then found by minimizing
the total energy.  This allows us to make predictions for the
Wilson coefficient of (\ref{effop}) in terms of the fundamental
parameters of the theory.  All of the nonperturbative physics is
encoded in the dimensionless ratio
\be
 	\zeta \equiv{|\psi(0)|^2\over m_\Phi^3}\,,
\label{zeq}
\ee 
It depends only on $\Lambda_\hc/M$, where $M$ is the common mass scale
for the new particles that we have assumed for simplicity. As shown
in fig.\ \ref{zetaf}, $\zeta$
is always small and is maximized near $\zeta \cong 0.004$ for 
$M\sim 2.5\Lambda_\hc$.  As benchmark values we will adopt
\be
	M = 1\,{\rm TeV},\quad \Lambda_\hc = 400\,{\rm GeV}
\label{bmark}
\ee

\section{Flavor Physics}

\begin{figure}[t]
\hspace{-0.4cm}
\centerline{
\includegraphics[width=0.5\hsize]{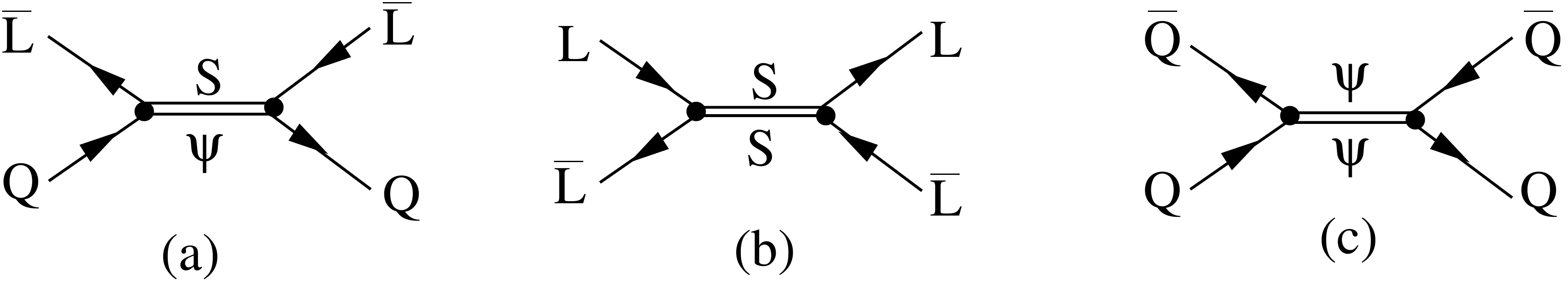}}
\caption{Flavor changing neutral currents mediated by the three
different kinds of bound states.}
\label{tree}
\end{figure}

We can fit the anomalies in $B\to K\ell\bar\ell$ decays by imposing
\be
	|\lambda_2^2\,\tilde\lambda_2\,\tilde\lambda_3|
	\cong 0.3\left(M\over{\rm TeV}\right)^2 
\left(3\over N_\hc\right)\,.
\label{fiteq2}
\ee
hence the relevant couplings can be reasonably small.  However it is
not trivial to find value that satisfy other flavor constraints.
This is because analogous exchanges of $\bar\Psi\Psi$ bound states 
give rise to meson-antimeson mixing, as illustrated in fig.\ 
\ref{tree}(c).  Especially for $B_s$ mixing, the same combination of
quark couplings $\tilde\lambda_2\tilde\lambda_3$ as in (\ref{fiteq2})
is relevant.  To keep them sufficiently small, we must take
$\lambda_2$ in (\ref{fiteq2}) to be sizable.   An example of 
values that can satisfy all constraints is
\bea
	\tilde\lambda_1 &=& -0.01,\quad
	\tilde\lambda_2 = 0.1,\quad
	\tilde\lambda_3 = 0.66,\quad % 0.61,\quad
	\lambda_2 = 2.1\nn \\
	(\tilde\lambda_1' &=& 0.014,\quad
	\tilde\lambda_2' = 0.13,\quad
	\tilde\lambda_3' = 0.66)\,.
\label{params}
\eea
The predicted values of products of couplings relevant to mixing
of the neutral mesons is shown in table \ref{tab2}.
We choose to saturate the $B_s$ mixing constraint 
\cite{Arnan:2016cpy}.

\begin{table}[b]
\centering
\begin{tabular}{|c|c|c|c|}
\hline
meson & quantity &$\hbox{ upper limit}\atop\hbox{ (units $M$/TeV)}$ 
& $\hbox{fiducial value}\atop\hbox{ (units $M$/TeV)}$\\
\hline
$K^0$ & $|\tilde\lambda_1\tilde\lambda_2|$ & $1.3\times
10^{-3}$ & $1\times 10^{-3}$ \\
$D^0$ & $|\tilde\lambda_1'\tilde\lambda_2'|$ & $2\times 10^{-3}$ & 
$7\times 10^{-4}$\\
$B^0$ & $|\tilde\lambda_1\tilde\lambda_3|$ & $0.026$ & $0.0066$\\
$B_s^0$ & $|\tilde\lambda_2\tilde\lambda_3|$ & $0.066$ & $0.066$\\
\hline
\end{tabular}
\caption{Predicted and limiting values of products of couplings
determining neutral meson mixing, assuming eq.\ (\ref{params}).}
\label{tab2}
\end{table}

In addition to meson mixing, there are radiative FCNCs like
$b\to s\gamma$, coming from transition magnetic moments between
heavy bound state quark partners $\Psi\phi$ and the SM quarks.
Because there is mass mixing induced by the interaction (\ref{Lint}) between these states,
in the mass basis the transition moment between heavy quark partners
and SM quarks induces transition moments between different flavors of
SM quarks, notably $b$ and $s$.  However the amplitude turns out to be
well below the current limit.

The previous processes have counterparts involving leptons, from 
fig.\ \ref{tree}(b).  They can be avoided by assuming
$\lambda_1=\lambda_3 = 0$ (the couplings to first and third generation
leptons), which is radiatively stable since to generate them from 
$\lambda_2$ at one loop requires a neutrino mass insertion.  But in
general one finds upper bounds on $\lambda_1$ and $\lambda_3$ from
$\mu\to 3e$, $\tau\to 3\mu$, and radiative transitions.  The most
stringent constraint arises from $\mu\to e\gamma$ and $\tau\to
\mu\gamma$, 
\be
	|\lambda_1| \lesssim 7.5\times 10^{-4},\quad |\lambda_3|
\lesssim 0.56\,,
\ee
The new contribution to $(g-2)_\mu$ is much smaller (by a factor of 300)
than needed to explain the outstanding discrepancy.

\section{Composite dark matter}
The new $S$ particle is neutral under SM interactions, and stable
by virtue of the accidental $Z_2$ symmetry, if it is the lightest
of the new particles.  The
baryon-like bound state $\Sigma = S^{N_\hc}$ is therefore a stable dark matter
(DM) candidate.   The nonrelativistic potential model predicts its mass to
be several TeV, given (\ref{bmark}).  Previous studies of composite
baryon-like DM in this mass and coupling range show that its thermal
relic density is highly suppressed by annihilations to hypergluons at
temperature above the confinement scale 
\cite{Mitridate:2017oky,Cline:2016nab}.  We must therefore assume
there exists some mechanism for generating an asymmetry in its number
density, which is conserved in our model (hyperbaryon number can be
consistently assigned to all the new particles).  

\begin{figure}[t]
\hspace{-0.4cm}
\centerline{
\raisebox{2cm}{\includegraphics[width=0.25\hsize]{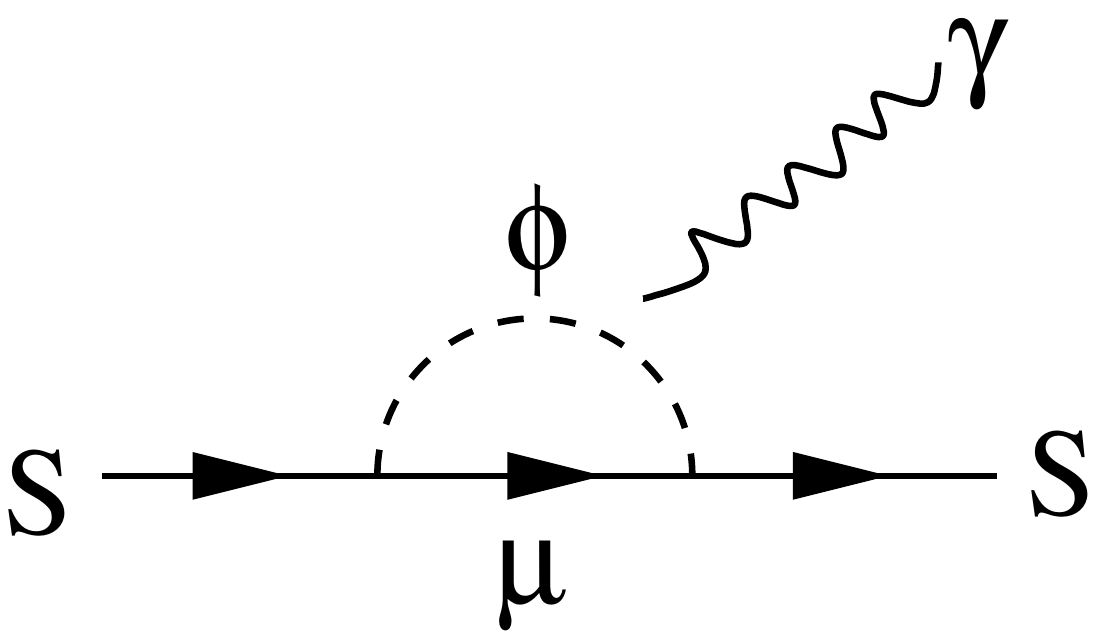}}
\hfil 
\includegraphics[width=0.5\hsize]{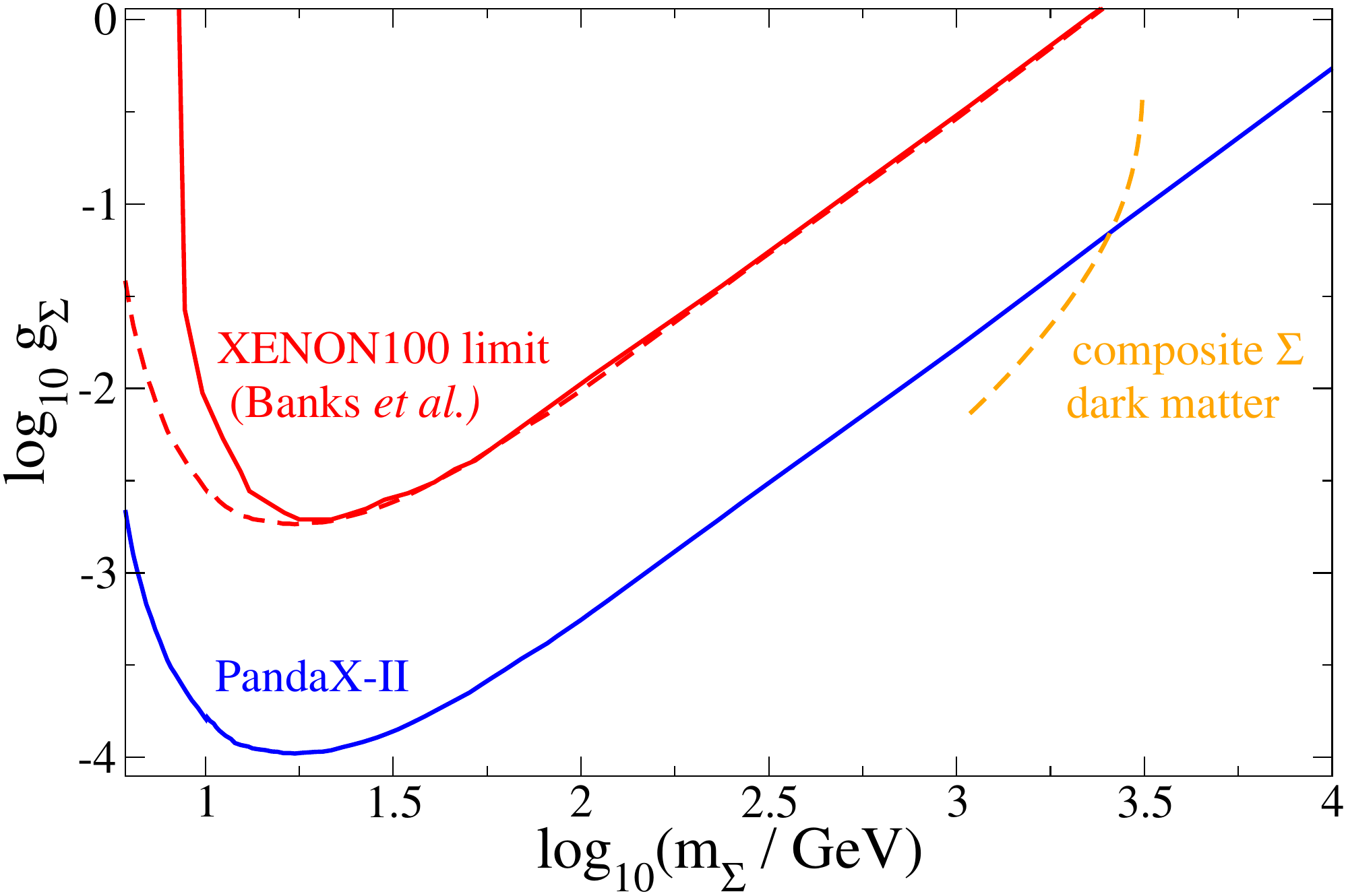}}
\caption{Left: diagram generating a magnetic moment for the $S$
fermion.  Right: direct detectioon constraint on the dark matter
$S^{N_\hc}$ gyromagnetic ratio.}
\label{DD}
\end{figure}

The model is strongly constrained by direct searches for the dark
matter, if $N_\hc$ is odd.  The $S$ fermion gets a magnetic moment
at one loop from the diagram in fig.\ \ref{DD}(a), 
\be
\mu_S = {e|\lambda_2|^2 m_S \over 32\pi^2\,m_\phi^2 }f(R)\,,
\ee
where $R \equiv m_S^2/m_\phi^2$ and the loop function $f(R)\sim 1$.
If $N_\hc$ is odd, $\Sigma$ is fermionic and inherits a magnetic
moment from its constituents of order $N_\hc \mu_S$.  
Updating older constraints on dark matter with a magnetic moment
\cite{Banks:2010eh}, we obtain fig.\ \ref{DD}(b) where the predicted
curve (dashed line) is parametrized by $m_S$.  Since $m_\Sigma$
varies rather weakly with $m_S$, due to the large contribution to its
mass from the hypergluons, the curve is steep as a function of 
$m_\Sigma$.  It is only below current limits for $m_S \lesssim 800\,$
GeV.

\section{LHC constraints}

\begin{figure}[t]
\hspace{-0.4cm}
\centerline{
\raisebox{0.8cm}{\includegraphics[width=0.4\hsize]{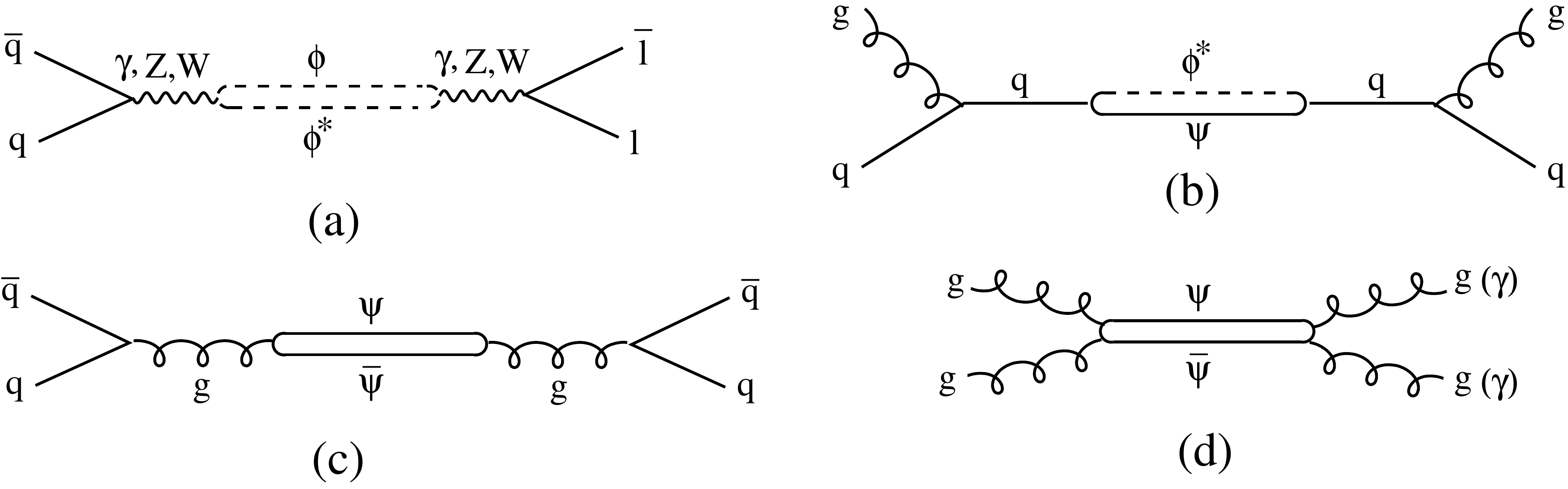}}
\hfil
\includegraphics[width=0.55\hsize]{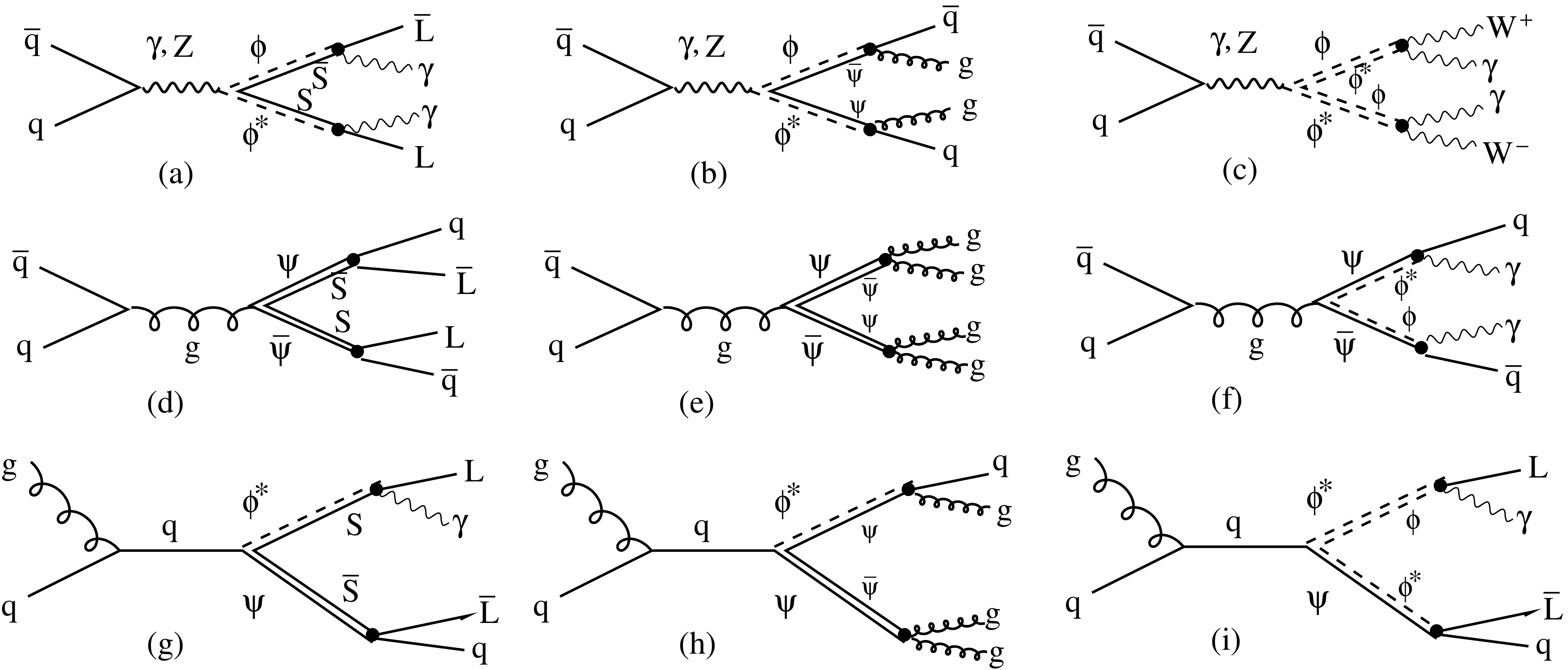}}
\caption{Left: Resonant production of  HC bound states
leading to dileptons, dijets  or diphotons; right: pair production
of bound states.}
\label{lhc-diagrams}
\end{figure}

Bound states can be produced resonantly at a hadron collider through
the processes shown in fig.\ \ref{lhc-diagrams}(left).  The parton level cross
sections can be computed in analogy to those for producing QCD bound
states like $J/\Psi$ at an electron collider.  For example the cross
section to produce the vector meson $\rho_\Psi = \bar\Psi\Psi$ 
from $q\bar q$ is
\be
	\sigma(q\bar q\to \rho_\Psi) = N_\hc{64\pi^3 \alpha_s^2 |\psi(0)|^2\over 
	3\,
	m_{\rho_\Psi}^3}\,\delta(s-m_B^2)
\ee
whose nonperturbative component resides in 
 the same ratio $\zeta$ as in (\ref{zeq}), given
that we have approximated all the bound state masses and wave
functions as being approximately the same.  Thus we can predict the
cross sections for these processes at LHC with no extra freedom from
adjusting parameters.

\begin{figure}[t]
\hspace{-0.4cm}
\centerline{
\includegraphics[width=0.45\hsize]{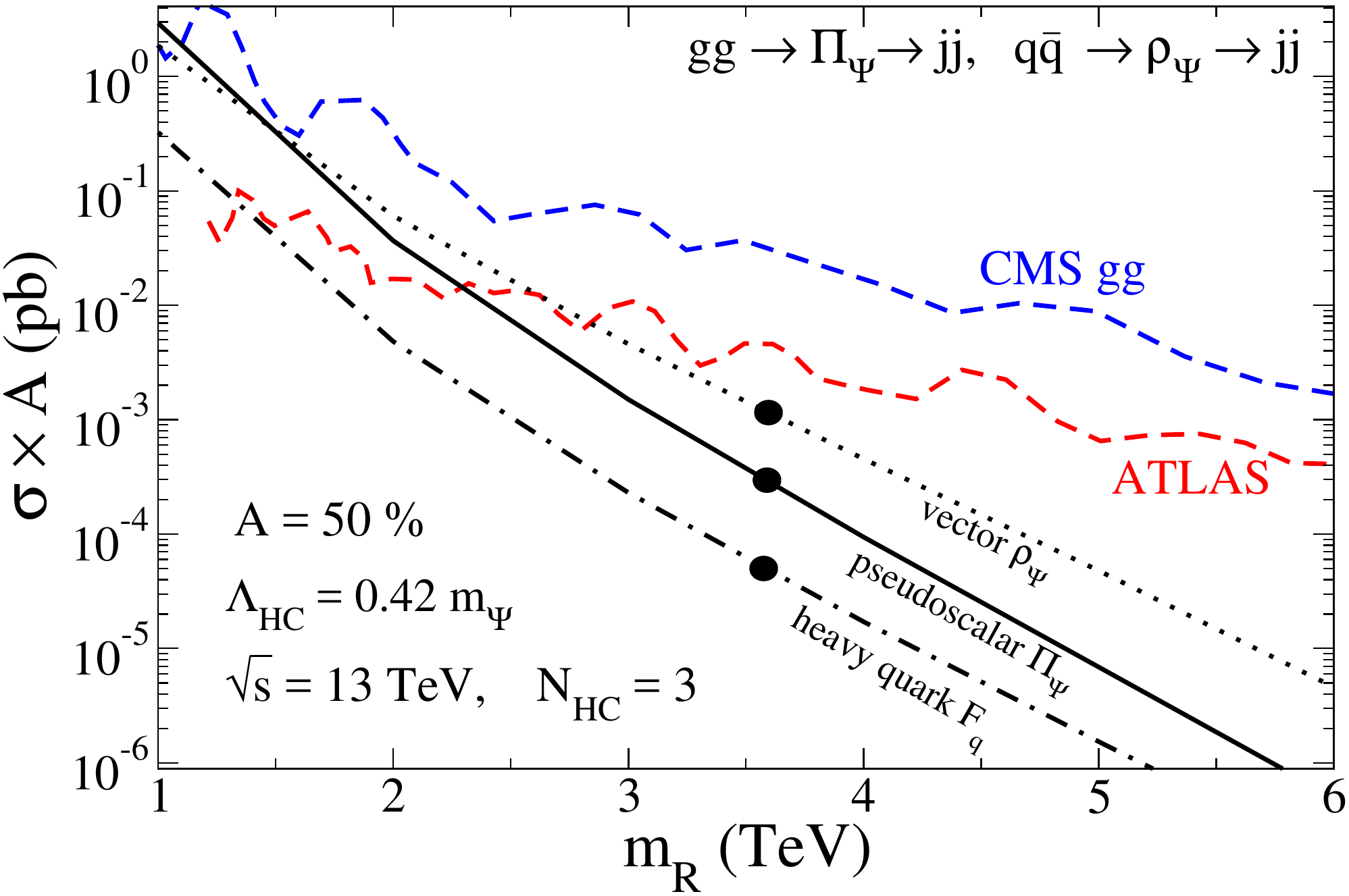}\hfil
\includegraphics[width=0.49\hsize]{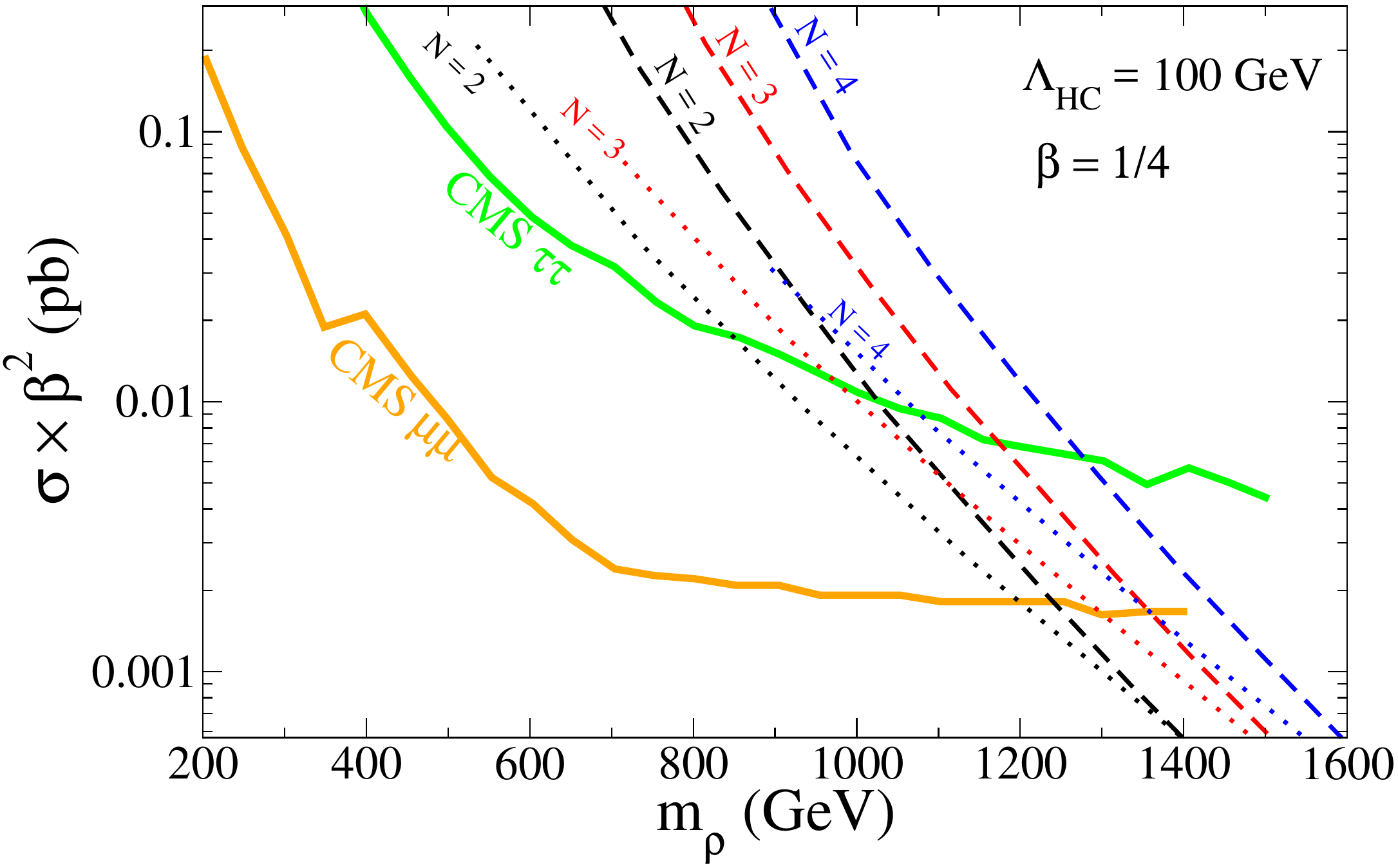}}
\caption{Left: LHC dijet limits; right: leptoquark search limits}
\label{lhc-limits}
\end{figure}

The processes in fig.\ \ref{lhc-diagrams}(left) produce dileptons, dijets or
diphotons.  ATLAS and CMS dijet constraints turn out to give the
most stringent limits on the model \cite{Aaboud:2017yvp,CMS:2017xrr},
shown in fig. \ref{lhc-limits}(left).  The $\rho_\Psi$ vector meson mass
must exceed 2.8\,TeV, which does not yet rule out our fiducial
model where all the resonances have mass $\cong 3.6\,$TeV.  The other
resonant states give rise to weaker limits.

It is also possible to pair-produce the bound states by open
production of the hyperquarks, followed by hadronization in the
SU($N_\hc$) sector, as shown in fig.\ \ref{lhc-diagrams}(right).  This requires more energy and leads to
large phase-space suppression of the cross sections.  If
$m_S\ll\Lambda_hc$ then this effect is mitigated for the states
containing $S$, including the leptoquarks.  Then leptoquark searches 
can be used to constrain the model, where final states with two
leptons and two jets are scrutinized 
\cite{CMS:2016qhm,Sirunyan:2017yrk}.  The limits, shown in fig.\
\ref{lhc-limits}(right) along with our model predictions for $m_S=0$,
are less constraining than those from the resonant production
searches.

\section{Conclusions}
Ours is not the first model of composite leptoquarks that has been
proposed to account for the $B\to K\ell\bar\ell$ decay anomalies, but 
we believe it is considerably simpler than others
\cite{Gripaios:2015gra,Barbieri:2016las,Matsuzaki:2017bpp,
Buttazzo:2017ixm}.  It is tightly constrained by FCNC processes,
namely meson mixing, and LHC searches for resonant production of
bound states of the heavy quark-like constituents.  It has the virtue
of providing a composite dark matter candidate $\Sigma$, that is also
challenged by current direct searches if $\Sigma$ contains an odd
number of constituents (hence is fermionic).  More accurate
predictions of the model could be obtained by a lattice study of the
SU($N_\hc$) 
bound state properties, that we have estimated in a rough manner.

\end{document}